\documentclass[prl,aps,superscriptaddress]{revtex4}

\usepackage{amsmath}
\usepackage{epsfig}
\begin{document}

\title{Observation of the skin-depth effect on the Casimir force between metallic surfaces}
\author{Mariangela Lisanti}
\affiliation{Harvard University, Department of Physics, Cambridge, MA 02138}
\author{Davide Iannuzzi}
\affiliation{Harvard University, Division of Engineering and Applied Sciences, Cambridge, MA 02138}
\author{Federico Capasso}
\email{capasso@deas.harvard.edu}
\affiliation{Harvard University, Division of Engineering and Applied Sciences, Cambridge, MA 02138}

\begin{abstract}
We have performed comparative measurements of the Casimir force between a metallic plate and a transparent sphere coated with metallic films of different thicknesses. We have observed that, if the thickness of the coating is less than the skin-depth of the electromagnetic modes that mostly contribute to the interaction, the force is significantly smaller than that measured with a thick bulk-like film. Our results provide the first direct evidence of the skin-depth effect on the Casimir force between metallic surfaces.

\end{abstract}

\maketitle

The long range attractive force between electrically neutral metallic plates \cite{casimir} or, more generally, dielectric bodies \cite{lifshitz}, known as the Casimir effect, has witnessed renewed experimental interest since the high precision measurements by S. K. Lamoreaux \cite{exp1}. All recent Casimir force experiments have been performed using surfaces covered with a thick metallic layer \cite{exp1,exp2,exp3,exp4,exp5,exp6,exp7,exp8}. In this case, the dielectric function of the film can be considered equal to the tabulated value for the corresponding bulk metal, and the expected force can be calculated with high accuracy. The comparison of the theoretical predictions with experimental results has allowed new limits to be set on the existence of extra-gravitational forces at short distances and has contributed to discussions concerning temperature corrections to the Casimir effect (see, for example, reference \cite{discussion}). 

On the other hand, the use of much thinner metallic coatings over transparent dielectrics should reveal an interesting phenomenon. At sub-micron distances, the Casimir force critically depends on the reflectivity of the interacting surfaces for wavelengths in the ultraviolet to far-infrared \cite{lifshitz,iannuzzi}. The attraction between transparent materials is expected to be smaller than that between highly reflective mirrors as a result of a less effective confinement of electromagnetic modes inside the optical cavity defined by the surfaces. A thin metallic film can be transparent to electromagnetic waves that would otherwise be reflected by bulk metal. In fact, when its thickness is much less than the skin-depth, most of the light passes through the film. Consequently, the Casimir force between metallic films should be significantly reduced when its thickness is less than the skin-depth at ultraviolet to infrared wavelengths. For most common metals, this condition is reached when the thickness of the layer is $\simeq 100$ \AA. 

In this letter, we present the first direct evidence of the skin-depth effect on the Casimir force between two dielectrics coated with metallic thin films. We have measured the Casimir force between a thick metal and a polystyrene sphere covered with a $\simeq 100$ \AA \medspace metallic film. The results are compared to those obtained after evaporating a thicker layer of metal ($\simeq 2000$ \AA) onto the same sphere. Our experiment shows that the Casimir attraction is significantly smaller when the sphere is coated with the thin film. This finding is confirmed by calculations.

Our experimental apparatus (see Fig. \ref{fig_setup}), which resembles the one described in reference \cite{exp6}, is designed to measure the force between a sphere and a plate at sub-micron distances with a force sensitivity on the order of $10$ pN. The measurement is carried out by positioning the sphere on top of a micro-machined torsional balance (MTB), and measuring the rotation angle of the balance induced by the Casimir attraction with the sphere as a function of the separation of the surfaces. 

The MTB is similar to a microscopic seesaw. Two thin torsional rods keep a gold-coated polysilicon plate (500 $\mu$m $\times$ 500 $\mu$m) suspended over two polysilicon electrodes symmetrically located on each side of the pivot axis. The capacitance between the top plate and each bottom electrode depends on the tilting angle $\theta$. When an external force $F$ induces a rotation of the top plate, one of the two capacitances increases by $\delta C\propto \theta \propto F$, while the other decreases by the same amount. An electronic circuit allows measurements of $\delta C$ with a sensitivity on the order of $10^{-6}$ pF, corresponding to $\theta\simeq 10^{-7}$ rad. Because the spring constant of the seesaw $k_s$ is about $10^{-8}$ Nm/rad, the sensitivity in the torque measurement is approximately equal to $k_s\theta\simeq 10^{-15}$ Nm, which corresponds to a force of $10$ pN in our experiment \cite{iannuzzi}.

The MTB is glued to a chip package and mounted inside a chamber that can be pumped down to $\simeq 10^{-3}$ mTorr. A 100 $\mu$m radius polystyrene sphere, mounted on the end of a rigid support and coated with a metallic layer, is clamped  to a manipulator that can bring the sphere close to the top plate of the MTB and controls the distance between the two surfaces. The manipulator consists of a triaxial stage for rough positioning and a piezoelectric translator (calibrated with an optical profiler) for fine tuning of the distance (see Fig. \ref{fig_setup}). 

To measure the Casimir force as a function of distance, we have followed the method described in reference \cite{iannuzzi}. After the chamber is evacuated, the piezoelectric stage is extended towards the MTB to reduce the separation between the sphere and the plate until the distance is only a few nanometers larger than the \emph{jump-to-contact} point (i.e., the distance at which the restoring torque of the seesaw is not sufficient to overcome the external torque induced by the Casimir force, causing the plate to come into contact with the sphere). The output of the capacitance bridge $A$ is then recorded as a function of the voltage applied to the sphere $V_{bias}$, which is scanned a few hundred millivolts around the so-called \emph{residual voltage} $V_0$ ($\simeq 200$ mV), i.e., the electrostatic potential drop arising from the difference of the work functions of the two films \cite{van} plus the potential difference generated by the metallic contacts of the electronics \cite{exp1}. The read-out system is designed so that $A$ is proportional to $\delta C$ and, therefore, to $F$:

\begin{equation}
A=c_1F=c_1\epsilon_0\pi R \frac{(V_{bias}+V_0)^2}{(d_0-d_{pz})}+c_1|F_C|
\label{eq_for_a}
\end{equation}

\noindent where $\epsilon_0$ is the permittivity of vacuum, $R$ is the radius of the sphere, $d_{pz}$ is the extension of the piezoelectric stage, $d_0$ is the distance between the sphere and the plate when the piezoelectric stage is not extended, and $F_C$ is the Casimir force. The distance $d$ between the sphere and the plate is given by $d_0-d_{pz}$; while the calibration of the piezoelectric stage provides $d_{pz}$ with a precision of less than 1 nm, $d_0$ is \emph{a priori} unknown and must be determined independently for an accurate comparison of experiment with theory \cite{iann_QFEXT}. In addition, it is worth stressing that $c_1$ is also unknown at this point, because the MTB has not yet been calibrated.

The measurement of $A$ as a function of $V_{bias}$ is then repeated for different values of $d$, which is changed by sequentially retracting the piezoelectric stage by a few nanometers.

For each value of $d$, data are interpolated with a generic quadratic equation $y=\alpha(x+x_0)^2+\beta$, where $\alpha$, $\beta$, and $x_0$ are free parameters. Note that

\begin{equation}
\alpha=c_1\frac{\epsilon_0 \pi R}{(d_0-d_{pz})}
\label{eq_for_alpha}
\end{equation}

\noindent By fitting $\alpha$ as a function of $d_{pz}$, it is thus possible to determine $d_0$ and $c_1$. Once $c_1$ is known, $F_C$ can be calculated by means of

\begin{equation}
|F_C|=\frac{\beta}{c_1}
\label{eq_for_fc}
\end{equation}

\noindent Because $d_0$ has also been determined, one can finally plot $F_C$ as a function of the distance between the sphere and the plate, $d=d_0-d_{pz}$.

Demonstrating the skin-depth effect requires careful control of the films' thickness and surface roughness. The sphere was glued to its support and subsequently coated with a $29\pm 2$ \AA \medspace titanium adhesion layer and a $92\pm 3$ \AA \medspace film of palladium. The thickness of the titanium layer and of the palladium film were measured by Rutherford Back Scattering on a silicon slice that was evaporated in close proximity to the sphere. After evaporation, the sphere was imaged with an optical profiler to determine its roughness, and mounted inside our experimental apparatus. After completion of the Casimir force measurements, the sphere was removed from the experimental apparatus, coated with an additional 2000 \AA \medspace of palladium, analyzed with the optical profiler, and mounted back inside the vacuum chamber for another set of measurements. It is important to stress that the surface roughness measured before and after the deposition of the thicker palladium layer was the same within a few percent. 

In Fig. \ref{fig_res}, we compare the results of the thin film measurements with those obtained after the evaporation of the thick layer of palladium. We repeated the measurement 20 times for both the thin and thick films. 

Our results clearly demonstrate the skin-depth effect on the Casimir force. The force measured with the thin film of palladium is in fact smaller than that observed after the evaporation of the thicker film. Measurements were repeated with a similar sphere: the results confirmed the skin-depth effect. To rule out possible spurious effects, we have compared our data with a theoretical calculation.

The Casimir force between a sphere and a plate can be calculated according to the well known Lifshitz equation \cite{lifshitz}:

\begin{equation}
F^{(L)}_C(d)=\frac{\hbar}{2\pi c^2} R \int_{0}^{\infty}d\xi\int_{1}^{\infty}dp \epsilon_3 p\xi^2\Big\{ \log\Big[1-\Delta^{(1)}_{31}\Delta^{(1)}_{32}e^{-x} \Big]+  \log\Big[1-\Delta^{(2)}_{31}
\Delta^{(2)}_{32}e^{-x}\Big]\Big\}
\label{eq_lif}
\end{equation}

\begin{equation}
\Delta^{(1)}_{jk}=\frac{s_k\epsilon_j-s_j\epsilon_k}{s_k\epsilon_j
+s_j\epsilon_k} \qquad \Delta^{(2)}_{jk}=\frac{s_k-s_j}{s_k+s_j}
\label{deltas}
\end{equation}

\begin{equation}
x=\frac{2d\sqrt{\epsilon_3}\xi p}{c} \qquad
s_k=\sqrt{p^2-1+\frac{\epsilon_k}{\epsilon_3}} \qquad 
\label{s_and_epsilon}
\end{equation}

\noindent where $\hbar$ and $c$ are the usual fundamental constants, and $\epsilon_1$, $\epsilon_2$, $\epsilon_3$ are the dielectric functions of the sphere, the plate, and the intervening medium, respectively, evaluated at imaginary frequencies $i\xi$

\begin{equation}
\epsilon_k(i \xi)=1+\frac{2}{\pi}\int_0^\infty \frac{x \cdot \textrm{Im}\epsilon_k(x)}{x^2+\xi^2}dx
\label{epsi}
\end{equation}

\noindent If the sphere is covered with an adhesion layer of thickness $t_4$ plus a coating film of thickness $t_5$, the force is still given by equation \ref{eq_lif} with $\Delta^{(1,2)}_{31}$ replaced with \cite{ninham}:

\begin{equation}
\Delta^{(1,2)}_{31} \rightarrow \frac{\Delta^{(1,2)}_{35}+
\Delta^{*(1,2)}_{51}e^{-\frac{xt_5s_5}{pd}}}
{1+\Delta^{(1,2)}_{35}
\Delta^{*(1,2)}_{51}e^{-\frac{xt_5s_5}{pd}}}
\label{sub_1}
\end{equation}

\begin{equation}
\Delta^{*(1,2)}_{51}=\frac{\Delta^{(1,2)}_{54}+\Delta^{(1,2)}_{41}e^
{-\frac{xt_4s_4}{pd}}}{1+\Delta^{(1,2)}_{54}\Delta^{(1,2)}_{41}e^
{-\frac{xt_4s_4}{pd}}}
\label{sub_2}
\end{equation}

\noindent where the subscripts 4 and 5 refer to the adhesion layer and to the coating film, respectively, as shown in Fig. \ref{fig_setup}. Surface roughness further modifies the Casimir force. This correction can be calculated according to \cite{decca_prd}:

\begin{equation}
F_C^{(L,\rho)}(d)=\sum_{i,j} v^{(sp)}_iv^{(pl)}_jF_C^{(L)}(d-(\delta^{(sp)}_i+\delta^{(pl)}_j))
\label{rough_corr}
\end{equation}

\noindent where $v_i$ is the probability that the surface of the sphere (superscript $sp$) or of the plate (superscript $pl$) is displaced by an amount $\delta_i$ with respect to the ideally smooth surface.

In Fig. \ref{fig_elab}, we compare our data with the theoretical result. The dielectric function used in the calculation was obtained from references \cite{poly,palig1,palig2,drude}. The values of $v$ as a function of $\delta$ were extracted from 10$\mu$m $\times$ 10$\mu$m images obtained with an optical profiler. At close distance ($\lesssim 100$ nm), our data are smaller than the prediction, both for the thin and for the thick film. This disagreement most likely lies in the fact that, in the analysis of the data, we have neglected the decrease of surface separation induced by the rotation of the MTB's top plate \cite{exp6}: this modifies equation \ref{eq_for_a}, and, thus, equation \ref{eq_for_alpha}. Furthermore, in equation \ref{eq_for_a} we have neglected the effect of surface roughness on the electrostatic force, which might induce errors in the determination of $d_0$ and $c_1$. Finally, it is worth noting that the calculated force at short distances strongly depends on the roughness of the two surfaces (equation \ref{rough_corr}), with corrective factors that, in our case, are as large as $\simeq 25\%$. These corrections might give rise to relevant errors in the calculations \cite{note_on_rough,note_on_rough_2}.  

It is also important to stress that the calculation of the force for the thin metallic film is based on two approximations: (i) the dielectric function for the metallic layers (both titanium and palladium) is assumed to be equal to the one tabulated for bulk-materials and is considered independent of the wave vector \textbf{k}, and (ii) the model used to describe the dielectric function of polystyrene is limited to a simplified two-oscillator approximation \cite{poly}. These assumptions might lead to significant errors in the estimated force for the case of thin films \cite{note_on_thin1}. 

It is clear, however, that our data represent a direct evidence of the skin-depth effect on the Casimir force. We have demonstrated that the Casimir attraction between a metallic plate and a metallized dielectric sphere depends on the thickness of the metal layer deposited on the sphere. In particular, if the coating is thinner than the skin-depth relative to the modes that mostly contribute to the interaction, the force is significantly smaller than what is expected for a thick, bulk-like film. This result might suggest interesting solutions for micro- and nanomachinery applications because it provides a technique to decrease the Casimir attraction between two DC-conductive surfaces kept at sub-micron distances. 

This work was partially supported by NSEC (Nanoscale Science and Engineering Center), under NSF contract number PHY-0117795.

\pagebreak
\newpage

\newpage
\begin{figure}[t]
\caption{Sketch of the experimental set-up (not to scale). Insets: (a) Sketch of the working principle of the micro-machined torsional device; (b) Layout of our experimental configuration: 1 is the polystyrene sphere, 2 is the gold coated top plate of the micro-machined torsional device, 3 is vacuum, 4 is the titanium adhesion layer, and 5 is the palladium film. \label{fig_setup}}
\end{figure}

\begin{figure}[t]
\caption{Experimental results of the measurement of the force as a function of the separation between the interacting surfaces. Dots indicate data obtained with metallic thick films, open circles indicate those obtained with thin films on the same sphere. \label{fig_res}}
\end{figure}

\begin{figure}[t]
\caption{Comparison between experimental data and theory. Dots indicate data obtained with thick films, open circles indicate those obtained with thin films. Continuous and dashed lines represent theoretical predictions for thick and thin films, respectively. \label{fig_elab}}
\end{figure}


\begin{thebibliography}{99}
\bibitem{note_on_rough_2} Optical measurements do not allow us to distinguish topological details whose typical dimensions are smaller than $\simeq 500$ nm. We have thus imaged the surface of one of the spheres used for thick film measurement using an AFM. The rms value obtained over a $1\times 1$ $\mu$m$^2$ ($\simeq 10$ nm) is slightly smaller than what was obtained with the optical profiler ($\simeq 15$ nm) over a larger area. Therefore, ignoring topological details not accessible to optical profiler measurements does not affect the calculation of surface roughness corrections.
\end{thebibliography}
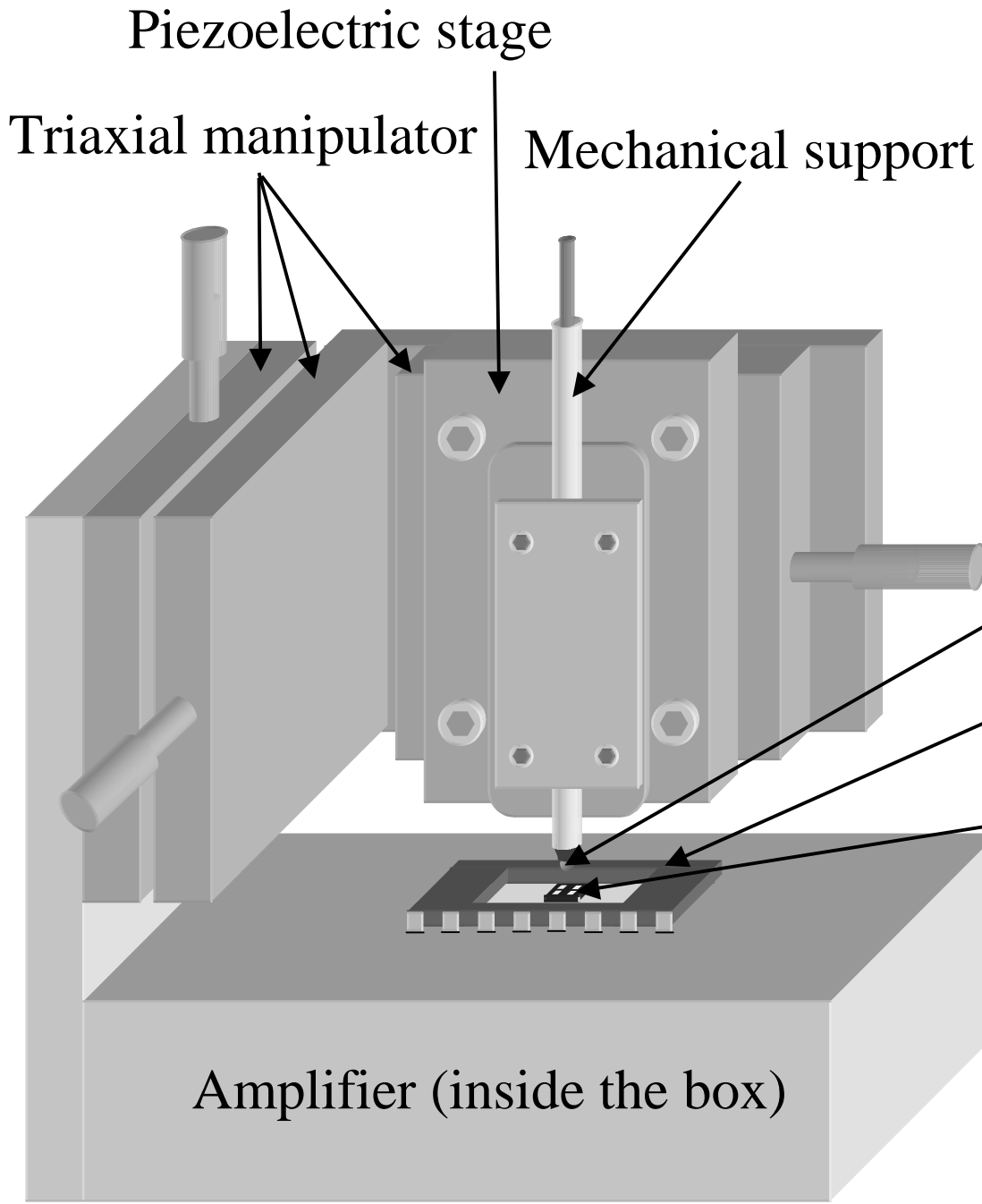
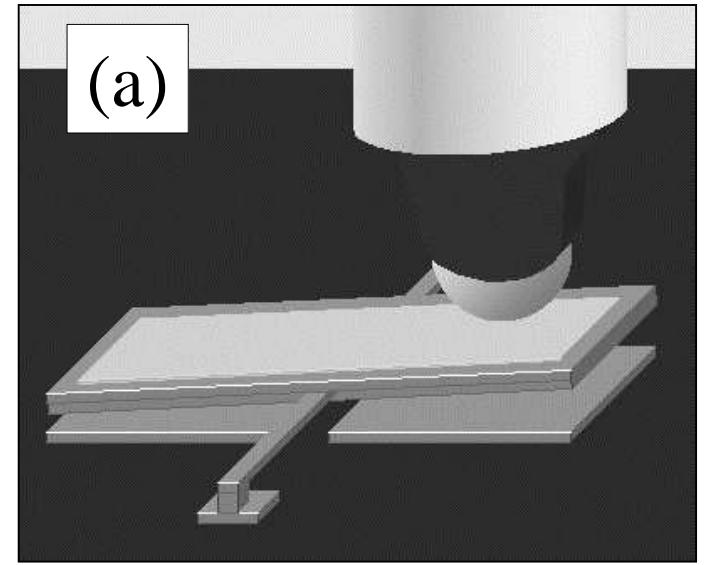
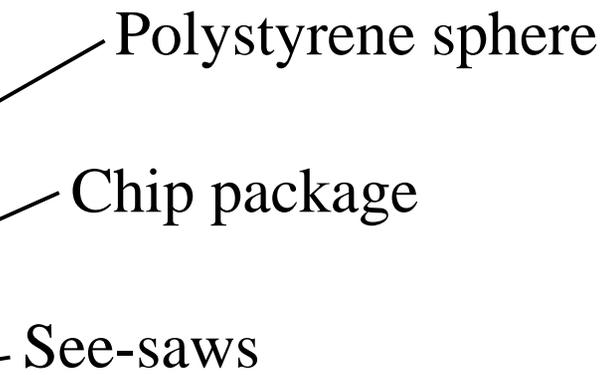
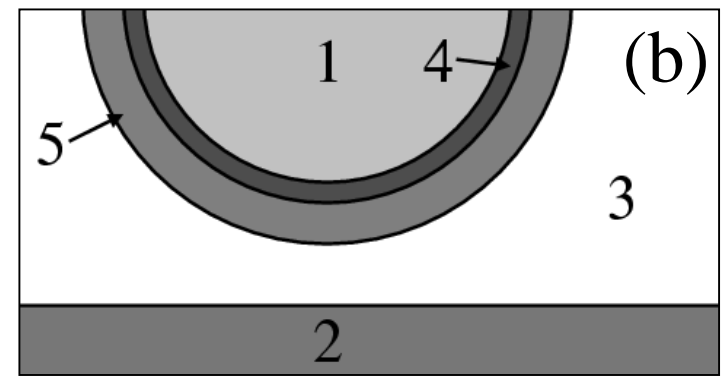
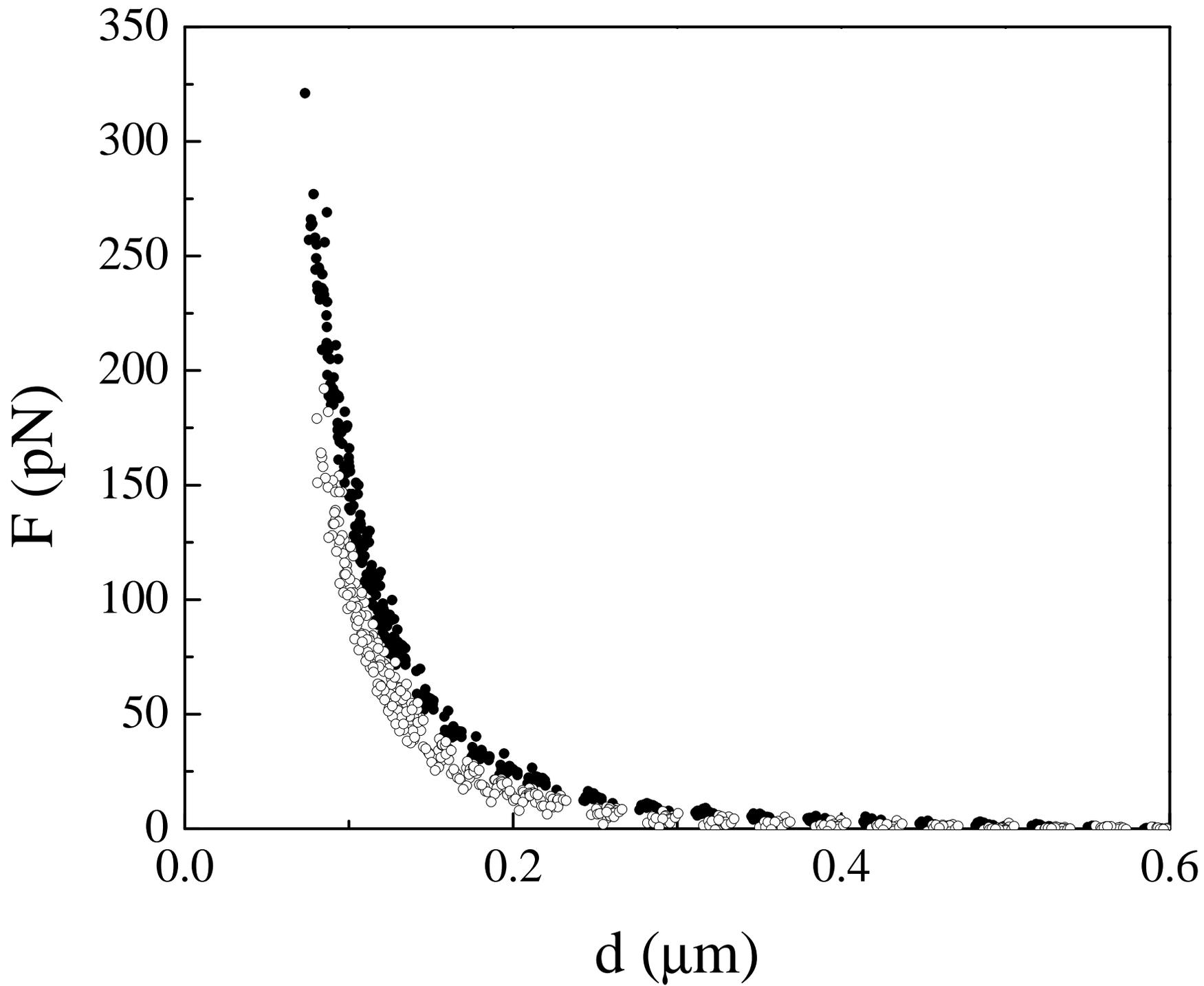
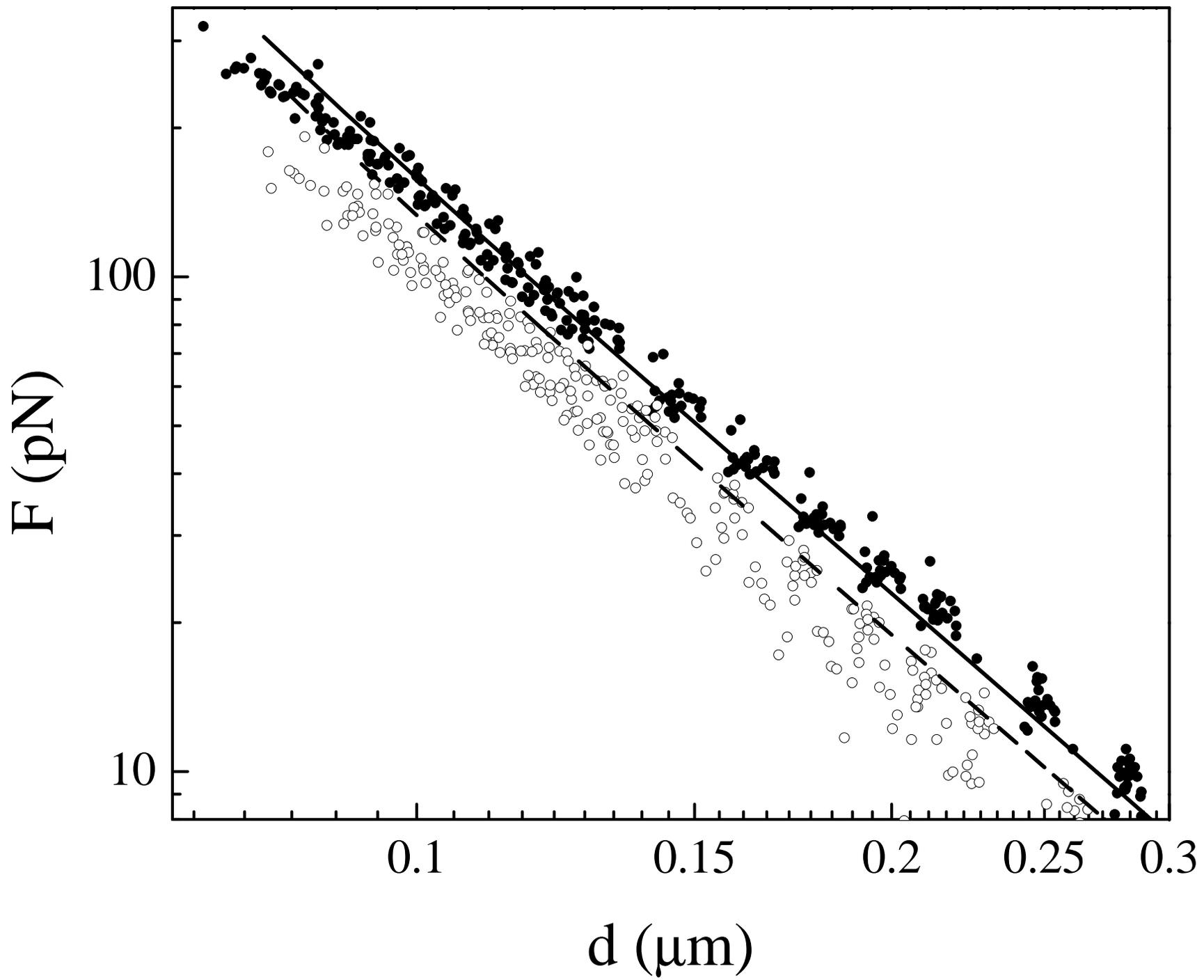
\end{document}